\documentstyle[prb,aps,twocolumn,floats,epsf]{revtex}
\begin{document}
\draft
\twocolumn[\hsize\textwidth\columnwidth\hsize\csname @twocolumnfalse\endcsname
\title{Non-monotonic magnetic field and density dependence of in-plane
magnetoresistance in dilute two-dimensional holes in GaAs/AlGaAs}
\author{H. Noh, Jongsoo Yoon,\cite{jsyoon} D. C. Tsui, and M. Shayegan}
\address{Department of Electrical Engineering, Princeton University,
Princeton, NJ 08544}
\date{\today}
\maketitle
\begin{abstract}
We studied low temperature ($T=50$mK) in-plane magnetoresistance of a 
dilute two-dimensional hole system in GaAs/AlGaAs heterostructure that 
exhibits an apparent metal-insulator transition. 
We found an anisotropic magnetoresistance, which changes dramatically 
at high in-plane fields ($B_{\parallel}\agt$5T) as the hole density is 
varied. At high densities where the system behaves 
metallic at $B_{\parallel}=0$, the transverse magnetoresistance is larger 
than the longitudinal magnetoresistance. With decreasing the hole density the 
difference becomes progressively smaller, and at densities near the 
"critical" density and lower, the longitudinal magnetoresistance becomes 
larger than the transverse magnetoresistance.
\end{abstract}
]

The scaling theory of localization\cite{localization} predicts that, 
in the absence of electron-electron interaction, 
all states in two dimensions (2D) are localized in the zero
temperature limit and that only an insulating behavior characterized by an
increasing resistivity ($\rho$) with decreasing temperature ($T$) is 
possible at low $T$. In contrast,
experimental observations of a "metallic" behavior, characterized by a
decreasing $\rho$ with decreasing $T$, and thus an apparent metal-insulator
transition (MIT) as charge carrier density is lowered, are reported on many
different low disorder dilute 2D systems at low temperatures.\cite{mit}
To date, despite the large number of experimental as well as
theoretical papers in the literature, there is still no consensus on either
the physics and the mechanism behind the metallic-like behavior or
the origin of this apparent MIT.

From the more recent experiments, two factors that strongly influence 
the charge transport close to this MIT in these low density high mobility 2D 
systems have emerged. First, disorder is found to play 
a significant role. A close examination of all such 2D systems reveals 
that the critical density ($n_{c}$ for electrons and $p_{c}$
for holes), the density above which the system shows the
metallic behavior, is 
lower in a system with higher carrier mobility.
The two most extensively studied systems are the electron inversion layer in
Si-MOSFET\cite{mosfet} and the 2D holes in the GaAs/AlGaAs 
heterostructure.\cite{p-gaas,yoon_0}
Typically, the 2D electrons in a high quality Si-MOSFET have a peak
mobility of $\sim2\times10^{4}$cm$^{2}$/Vs to $5\times10^{4}$cm$^{2}$/Vs, 
and $n_{c}\sim 1\times10^{11}$ cm$^{-2}$. In the extremely high mobility 
2D hole system (2DHS) in GaAs/AlGaAs, which has a comparable 
effective mass at low densities and a more than 10 times higher peak 
mobility of $\sim7\times10^{5}$cm$^{2}$/Vs, 
$p_{c}\sim 1\times10^{10}$ cm$^{-2}$. 
Both $n_{c}$ and $p_{c}$ are found to decrease with 
decreasing disorder in the 2D system,\cite{yoon_0} 
approaching the critical density
characterized by the dimensionless density parameter $r_{s}=37$, expected
for Wigner crystallization in 2D.\cite{tanatar}
	
The second factor is the effect of a magnetic field applied parallel to
the 2D plane. The parallel field ($B_{\parallel}$) is found to suppress 
the metallic behavior and also induce an unexpectedly large positive 
magnetoresistance (MR). In the case of the 2DHS in GaAs,\cite{yoon} this giant
MR follows $\sim exp(B_{\parallel}^{2})$ at low $B_{\parallel}$
and crosses into an $\sim exp(B_{\parallel})$ dependence at high 
$B_{\parallel}$ around a well defined characteristic field $B^{*}$,
which is hole density dependent. This behavior persists into the insulating
phase, though $B^{*}$ becomes density independent. Similarly strong
positive MR was observed earlier in the electron inversion
layer in Si-MOSFET's at low $B_{\parallel}$ though it is followed by 
saturation\cite{simonian} or in some cases a weak linear dependence 
at high $B_{\parallel}$.

Recently, Das Sarma and Hwang\cite{das-sarma} pointed out the 
importance of taking into account in the $B_{\parallel}$ transport the
finite thickness of the 2D electrons through which $B_{\parallel}$ can be
coupled into their orbital motion. They calculated the Boltzmann transport
in the presence of $B_{\parallel}$ assuming $\delta$-function random impurity
scattering and a harmonic confinement of the 2D carriers to the heterojunction
interface. They were able to produce a giant positive MR
showing $B_{\parallel}$ dependencies in qualitative agreement with 
experimental observations. In particular, their calculation showed almost
two orders of magnitude increase in resistance, when $B_{\parallel}$
was increased to 10T. Furthermore, they found a crossover from a low field
to high field behavior due to the change of the relative magnitude of the
cyclotron energy to the confinement energy with increasing field.
The difference between the $B_{\parallel}$ behavior seen in the Si inversion
layer and that in the 2DHS in GaAs in the high field regime was attributed
to the different impurity scattering mechanisms.
Most importantly for experiments, they predicted that this giant in-plane
MR is highly anisotropic: the transverse 
MR ($\rho_{\perp}$, the MR when 
$B_{\parallel} \perp I$, where $I$ is the current) can be many times 
larger than the longitudinal MR ($\rho_{\parallel}$, the
MR when $B_{\parallel} \parallel I$) for a moderately
high $B_{\parallel}$ in the 10T range.

In the course of a systematic study on the nature of the insulating
ground state in the extreme dilute limit of the 2DHS in GaAs, we have
investigated the $B_{\parallel}$ transport in one of the highest mobility low
density sample currently available as a function of the hole density ($p$) 
across the apparent MIT with $B_{\parallel}$ up to 14T at dilution refrigerator
temperatures down to 50mK. 
By exploring a regime (lower density, lower temperature) 
not accessible in previous in-plane MR anisotropy experiments, 
we find a behavior which cannot be understood with currently available models.
In addition to observing the expected behavior of a much larger 
MR in this regime, we find that the anisotropy, 
defined as the difference between $\rho_{\parallel}$ and $\rho_{\perp}$, 
changes sign in the high field region as the hole density is decreased.
For densities $p>2\times10^{10}$cm$^{-2}$,
$\rho_{\perp}$ is larger than $\rho_{\parallel}$ at high field
similar to the model calculation by Das Sarma and Hwang. 
However, the difference between $\rho_{\perp}$ and $\rho_{\parallel}$ 
becomes progressively smaller as the density is decreased,
and for densities $p<2\times10^{10}$cm$^{-2}$, $\rho_{\parallel}$ 
becomes larger than $\rho_{\perp}$. 
In the rest of this paper, we describe in more detail the experiment 
and discuss our data in light of more recent results from $^{3}$He temperature
experiments in the metallic regime.\cite{papadakis}
We also point out that the observed anisotropy is at variance with 
predictions of existing theories explaining the large in-plane
MR.

Our samples are made from a modulation doped GaAs/AlGaAs (311)A single 
interface heterostructure grown by molecular beam epitaxy. 
The 2DHS at the (311)A interface even at $B_{\parallel}=0$ exhibits a mobility 
anisotropy known to be due to
anisotropic surface morphology.\cite{311_aniso}
The mobility is high along the [\={2}33] crystallographic direction, 
and low along the [01\={1}] direction.
We made two Hall bar (rectangular) samples, one for each direction, 
from the same wafer.
For resistivity measurements, $I$ was always passed along the sample
length. Both samples also have back gates to change their density.
The peak mobilities at $T=50$mK for our ungated samples with a density of 
$p=4.2\times10^{10}$cm$^{-2}$ are $6.3\times10^{5}$cm$^{2}$/Vs 
along [\={2}33] and $4.2\times10^{5}$cm$^{2}$/Vs along [01\={1}].
These are among the highest mobilities achieved in samples of this 
type of single interface heterostructures. 
Our measurements were made down to $p=7.9\times10^{9}$cm$^{-2}$.
This density is about 40\% lower than that accessed by Papadakis 
{\it et al.} in their recent experiments using the GaAs/AlGaAs
(311)A quantum wells at $^{3}$He temperatures.\cite{papadakis}
It enabled us to measure the in-plane MR across the
zero-field MIT at $p_{c}=9.3\times10^{9}$cm$^{-2}$.
For a given density, we measured both 
$\rho_{\perp}(B_{\parallel})$ and $\rho_{\parallel}(B_{\parallel})$
by applying $B_{\parallel}$ up to 14T. 
The samples were mounted, in separate cooldowns,
on a rotating stage in a dilution refrigerator,
which could adjust the angle between the 2D plane and the direction of the
magnetic field.
They were mounted in such a way that either $B_{\parallel} \perp I$
for $\rho_{\perp}$ measurement or $B_{\parallel} \parallel I$ 
for $\rho_{\parallel}$ measurement, when the 2D plane is aligned parallel to
the magnetic field. 
Between the $\rho_{\perp}$ and $\rho_{\parallel}$ measurements,
the samples were thermally recycled, and
measurements were repeated for consistency check.

\begin{figure}[!t]
\begin{center}
\leavevmode
\hbox{%
\epsfxsize=3.25in
\epsffile{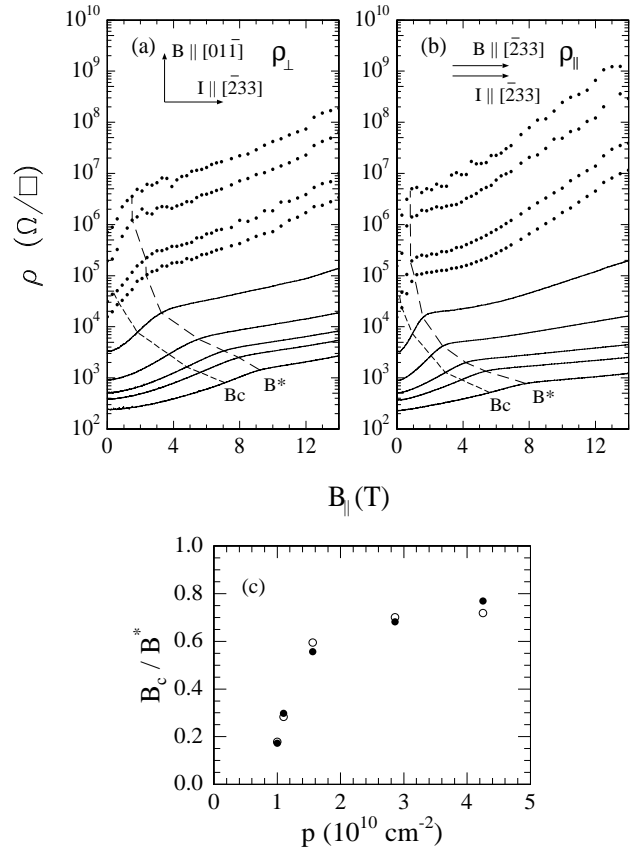}}
\end{center}
\caption{(a) $\rho_{\perp}$ and (b) $\rho_{\parallel}$
plotted as a function of $B_{\parallel}$ at densities 
$p=$ 4.25, 3.25, 2.86, 2.18, 1.56, 1.10, 1.00, 0.87, 0.79 
$\times10^{10}$cm$^{-2}$ (from bottom to top).
The critical density $p_{c}$, where the MIT occurs at zero-field, is
$p_{c}=0.93\times10^{10}$cm$^{-2}$. The characteristic field $B_{c}$, above 
which the metallic behavior is suppressed, is marked by the short dashed line,
and $B^{*}$, which divides the low field and the high field regions, is marked
by the long dashed line. The orientation of $B_{\parallel}$ and $I$ are shown
with respect to the crystal axes.
In both (a) and (b), the solid lines are data measured by low frequency 
lock-in technique and the solid circles are those measured by 
DC I-V measurement.
(c) $B_{c}/B^{*}$ plotted as a function of $p$. Solid (open) symbols 
are for $B_{\parallel} \perp I$ ($B_{\parallel} \parallel I$).}
\label{1}
\end{figure}

The in-plane MR, $\rho_{\perp}(B_{\parallel})$ and 
$\rho_{\parallel}(B_{\parallel})$, measured at 50mK for $I$ fixed in [\={2}33]
direction are
shown in Fig.~\ref{1} (a) and (b), respectively, in the density range 
from $4.25\times10^{10}$ cm$^{-2}$ to $0.79\times10^{10}$ cm$^{-2}$. 
The sample shows insulating behavior at $B_{\parallel}=0$ for
the two lowest densities (the top two traces) and metallic behavior for
the higher densities. Both $\rho_{\perp}$ and $\rho_{\parallel}$ 
show strong increase with increasing $B_{\parallel}$, 
about one order of magnitude for the highest density and
more than two ($\rho_{\perp}$) or three ($\rho_{\parallel}$) orders of 
magnitude for the lowest density at 14T. It is clear that the
MR is larger for lower densities. We should also note that
the MR in our sample is approximately ten times larger than that observed
by Papadakis {\it et al.}\cite{papadakis} at $T=0.3$K in the density range
where the two experiments overlap. 

For both $\rho_{\perp}$ and $\rho_{\parallel}$, according to their
dependence on $B_{\parallel}$, the $\rho$-$B_{\parallel}$ plane can
be divided into two regions: a low field region ($B_{\parallel}<B^{*}$) 
and a high field region ($B_{\parallel}>B^{*}$).
In the low field region, $\rho$ measured in both field orientations 
increases rapidly with increasing $B_{\parallel}$, following 
$\rho \sim exp(B_{\parallel}^{2})$. This rapid increase 
slows down as $B_{\parallel}$ is increased to the high field region, where
$\rho_{\perp}$ increases as $exp(B_{\parallel})$,
while $\rho_{\parallel}$ shows a behavior depending on the density.
For densities $p>2.18\times10^{10}$cm$^{-2}$, $\rho_{\parallel}$ also increases
as $exp(B_{\parallel})$, but for lower densities, it shows
a stronger dependence, which becomes more pronounced when the density is 
lowered. For both $B_{\parallel}$ orientations, the metallic behavior 
observed at $B_{\parallel}=0$ and $p>p_{c}$ is suppressed for 
$B_{\parallel}>B_{c}$ (the short dashed lines in Fig.~\ref{1} (a) and (b)).
Both $B_{c}$ and $B^{*}$ are smaller in Fig.~\ref{1}
(b) than in Fig.~\ref{1} (a). 
However, as shown in Fig.~\ref{1} (c), when the ratio of
$B_{c}$ to $B^{*}$ is plotted as a function of $p$, the two sets of
data are nearly identical. 
These observations result from a $B_{\parallel}$ induced spin subband
depopulation identified recently by Papadakis {\it et al.}\cite{papadakis} and
Tutuc {\it et al.}\cite{tutuc} from their detailed analyses of the
Shubnikov-de Haas oscillations observed at 0.3K. We shall later return to
this point.

\begin{figure}[!t]
\begin{center}
\leavevmode
\hbox{%
\epsfxsize=3in
\epsffile{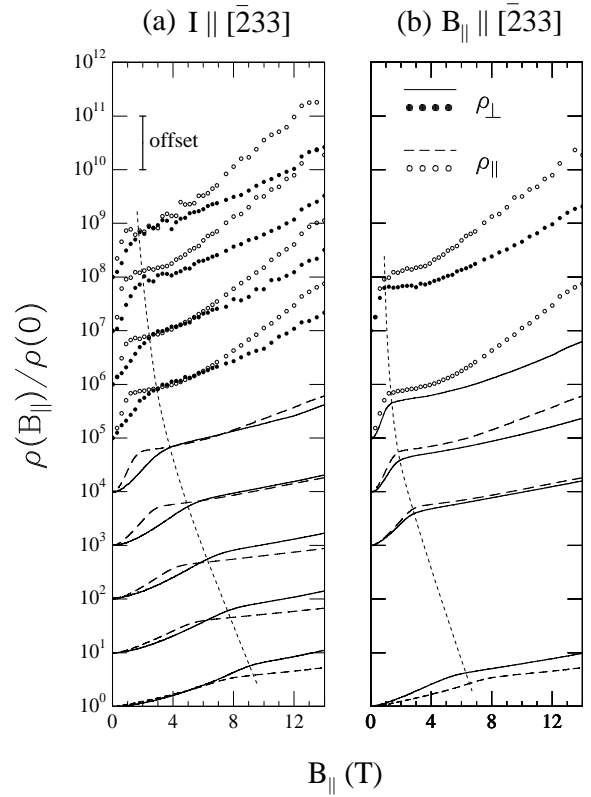}}
\end{center}
\caption{$\rho_{\perp}$ and $\rho_{\parallel}$ normalized by the 
zero-field resistivity $\rho(0)$ and plotted together for 
(a) $I$ fixed along [\={2}33] and (b) $B_{\parallel}$ fixed along [\={2}33].
The solid lines and solid circles are for $\rho_{\perp}$ and the dashed lines
and open circles are for $\rho_{\parallel}$.
The densities for each pair of traces in (a) from bottom to top is
$p=$ 4.25, 3.25, 2.86, 2.18, 1.56, 1.10, 1.00, 0.87, 0.79 
$\times10^{10}$cm$^{-2}$.
The data for each $p$ are offset by 10 as indicated 
by the scale bar for clarity. The densities in (b) from bottom to top is
$p=$ 4.25, 2.18, 1.56, 1.10, 0.87 $\times10^{10}$cm$^{-2}$ and the data
are offset in such a way that zero-field values match with those in (a)
for corresponding densities. The dotted lines are guide to the eye, 
representing the boundary between the low and high field regions.}
\label{2}
\end{figure}

In Fig.~\ref{2} (a), we plot both $\rho_{\perp}(B_{\parallel})$ and
$\rho_{\parallel}(B_{\parallel})$ from Fig.~\ref{1} normalized by 
the values at $B_{\parallel}=0$ in a single plot to show the anisotropy. 
Note that in Fig.~\ref{2} (a) both sets of data were taken with $I$
fixed along [\={2}33] and the direction of $B_{\parallel}$ was varied
from [01\={1}] to [\={2}33]
to measure $\rho_{\perp}$ and $\rho_{\parallel}$, respectively.
In Fig.~\ref{2} (b), we show additional data, taken with $I$ along [01\={1}]
and $B_{\parallel}$ along [\={2}33] ($\rho_{\perp}$), and compare them with
$\rho_{\parallel}$ of Fig.~\ref{2} (a), which is taken with both $I$ and
$B_{\parallel}$ along [\={2}33]. In other words, in Fig.~\ref{2} (b)
the direction of $B_{\parallel}$ is fixed along [\={2}33],
and $\rho_{\perp}$ and $\rho_{\parallel}$ data are shown for $I$ along
[01\={1}] and [\={2}33], respectively.
The data of Fig.~\ref{2} highlight the central finding of our paper:
{\it in the high field region, on the right of the dotted lines,
we observe an anisotropy in MR which reverses sign as
a function of density}. At high densities $\rho_{\perp}$ is larger
than $\rho_{\parallel}$, while at low densities $\rho_{\parallel}$ is
larger than $\rho_{\perp}$. 
This $\rho_{\parallel}>\rho_{\perp}$ anisotropy continues to grow across
the MIT, and for $p \alt p_{c}$, in the $B_{\parallel}=0$ insulating side, 
$\rho_{\parallel}$ is approximately an order of magnitude larger than 
$\rho_{\perp}$ at 14T. 
Note that this trend is seen regardless of
whether $\rho_{\perp}$ and $\rho_{\parallel}$ are measured by fixing the
direction of $I$ (Fig.~\ref{2} (a)) or $B_{\parallel}$ (Fig.~\ref{2} (b)),
suggesting that the relative orientation of $I$ and $B_{\parallel}$ 
is important in the high field region.
No change of anisotropy at high field with decreasing density was 
observed in earlier $T=0.3$K experiments,\cite{papadakis} where a 
relatively high density sample, showing only the metallic behavior at 
$B_{\parallel}=0$, was studied. 
There, $\rho_{\perp}$ was always larger than $\rho_{\parallel}$
even at $B_{\parallel}$ up to 16T.
Before we further discuss this surprising trend, it is useful to make a
few remarks about the MR data at lower $B_{\parallel}$,
on the left of the dotted lines in Fig.~\ref{2}.

As mentioned before, each of our MR traces has a characteristic
field, $B^{*}$. It has been recently shown that $B^{*}$ marks the field above
which the 2DHS becomes fully spin polarized.\cite{papadakis,tutuc,winkler}
Below $B^{*}$, two spin subbands with unequal populations are occupied
while, above $B^{*}$, there is only one occupied spin subband. The situation
is similar for 2D electron systems in Si-MOSFET's; there too a strong
MR is observed at low fields when two spin subbands are
populated, while the MR is essentially saturated once the
system is fully spin polarized.\cite{okamoto,vitkalov}
Calculations by Dolgopolov {\it et al.}\cite{dolgopolov} and 
Herbut\cite{herbut} attribute this low field MR to the
change of screening as the spin polarization changes.
In the case of 2DHS in GaAs (311)A, $B^{*}$ depends on the orientation of
the $B_{\parallel}$ with respect to the crystal axes. In particular, $B^{*}$
is smaller for $B_{\parallel}\parallel$[\={2}33] compared to
$B_{\parallel}\parallel$[01\={1}]. This is due to the intrinsic energy
band anisotropy of the GaAs (311)A 2DHS, as has been documented experimentally
and theoretically.\cite{papadakis,tutuc,winkler}
Our observation that $B^{*}$ in the $\rho_{\perp}$ data of Fig.~\ref{1} (a)
is larger than in the $\rho_{\parallel}$ data of Fig.~\ref{1} (b) is
consistent with these findings.
That the data plotted as $B_{c}/B^{*}$ for the two $B_{\parallel}$ orientations
in Fig.~\ref{1} (c) should closely agree is also to be expected if the
minority spin population is constant at $B_{c}$ as found by Tutuc 
{\it et al.}\cite{tutuc} and the spin polarization 
is linear in $B_{\parallel}$.

It is likely that the MR anisotropy observed at low 
fields in Fig.~\ref{2} (a) is a consequence of the difference 
in spin subband depopulation rate for different orientations of 
$B_{\parallel}$ with respect to the crystal axes. 
An alternative interpretation of the low field anisotropy may be based on
a recent theory by Chen {\it et al.}.\cite{chen}
These authors find that an interplay between the spin-orbit coupling and
the Zeeman splitting can lead to an anisotropic MR depending
on the relative direction between $I$ and $B_{\parallel}$.
The anisotropy is small at low and high fields, and exhibits a maximum at
an intermediate field when the Zeeman energy equals the spin-orbit induced
spin-splitting. At first sight, this seems qualitatively consistent with
the low field portion of Fig.~\ref{2} (a) where the difference between
$\rho_{\parallel}$ and $\rho_{\perp}$ goes through a maximum. However,
this trend is not observed in Fig.~\ref{2} (b) which also compares
$\rho_{\parallel}$ and $\rho_{\perp}$ but now for a fixed $B_{\parallel}$
direction. 
We conclude that the explanation of the anisotropic
MR observed in the low field region in Fig.~\ref{2} (a)
most likely lies in the anisotropic spin subband depopulation.\cite{papadakis}

We now return to a discussion of the high field data which are the main
subject of this paper. We first note that these data are in a field region
where only one spin subband is occupied and the system is fully spin
polarized. In this region, calculations such as those reported in Refs. 16 
and 17 predict that the MR would saturate. A saturation of
the MR above $B^{*}$ is indeed often seen in 2D electron
systems in Si-MOSFET's.\cite{okamoto,vitkalov} This is in contrast, however,
with the data shown in Fig.~\ref{2} where a large MR is
observed even at high fields beyond $B^{*}$. Calculations by Das Sarma
and Hwang,\cite{das-sarma} on the other hand, point to the coupling of
$B_{\parallel}$ to the orbital motion, and show that there can be a
significant MR when the magnetic length becomes comparable to
or smaller than the layer thickness. They also predict that the
MR should be anisotropic with $\rho_{\perp}>\rho_{\parallel}$.
While this trend is seen in our data at high densities (Fig.~\ref{2}),
the anisotropy is reversed at low densities.
Their calculations predict that with decreasing density, the anisotropy
($\rho_{\perp}>\rho_{\parallel}$) increases because of the increasing layer
thickness. This is also at odds with our experimental data, although
we should note that in our samples we expect the layer thickness to
decrease with decreasing density since we are using a back gate.
It is remarkable that the low density data indicate that at high fields
$\rho_{\parallel}>\rho_{\perp}$ regardless of whether the direction of
$I$ or $B_{\parallel}$ is kept fixed. This observation suggests that the
anomalous anisotropy we observe at low densities may be intrinsic to very
dilute 2D hole systems in GaAs.

Finally, Meir\cite{meir} explained the
large in-plane MR in GaAs as from inhomogeneous
distribution of charge carriers due to disorder.
In his model, the disorder potential induces puddles of 2D holes which
are interconnected by quantum point contacts (QPC), and the transport is 
dominated by the conduction through these QPC's.
An in-plane magnetic field increases the bottom of the 2D
subbands and effectively increases the conduction threshold
energy of the QPC's, giving rise to the large positive MR.
The MIT is a percolation transition which can be tuned by
density and by in-plane magnetic field because the conduction through the
QPC's depends on the relative value of the Fermi energy to the threshold
energy of the QPC's. However, no anisotropy is expected in this model, 
where the MR arises from the shift of the 2D subband energy by the 
presence of $B_{\parallel}$.

In summary, we have measured the in-plane MR $\rho_{\perp}$
and $\rho_{\parallel}$ of a very high mobility, dilute 2DHS in the
GaAs/AlGaAs heterostructure in the density range from 
$4.2\times10^{10}$cm$^{-2}$ to $7.9\times10^{9}$cm$^{-2}$, spanning across
the apparent zero-field MIT. In the low $B_{\parallel}$
region, the observed anisotropy is consistent with the spin subband
depopulation which depends on the orientation of $B_{\parallel}$ relative
to the crystal axes.
In the high field region, we observed a change in the sign of the 
anisotropy with decreasing density.
At high densities, $\rho_{\perp}>\rho_{\parallel}$. 
However, as the density is lowered, this anisotropy decreases
and $\rho_{\parallel}$ becomes equal to $\rho_{\perp}$
around $\sim 2\times10^{10}$cm$^{-2}$. For still lower
densities, $\rho_{\parallel}>\rho_{\perp}$, and the anisotropy changes sign
and also grows continuously across the MIT. These experimental observations
are not explicable with existing theoretical models for the in-plane MR. 

We would like to thank S. Das Sarma, E. H. Hwang, Y. Meir, P. Phillips,
and M. Hilke for fruitful discussions.
This work was supported by the NSF.

\end{document}